\documentclass[runningheads]{llncs}
\usepackage[T1]{fontenc}
\usepackage{graphicx}
\usepackage{amsmath}
\usepackage{amssymb}
\usepackage{booktabs}
\usepackage{algorithm}

\begin{document}

\title{Peak-DNA: Recurrent Morphology in Global Narrative Propagation}
\titlerunning{Peak-DNA}

\author{Joosung Kim \and Luyonghe Li \and Shanwen Wu}

\authorrunning{J. Kim et al.}

\institute{Graduate School of Informatics, Nagoya University\\
Furo-cho, Chikusa-ku, Nagoya, Aichi 464-8601, Japan\\
\email{cocobar@nagoya-u.jp}
}
\maketitle 

\begin{center}
\small\textit{Accepted at the 18th International Conference on Advances in
Social Networks Analysis and Mining (ASONAM 2026), Rabat, Morocco,
August 24--27, 2026.}
\end{center}

\begin{abstract}

Global narrative streams exhibit noisy, fragmented, and asynchronous amplification bursts, making direct waveform comparison unreliable for recurrence analysis. This paper introduces \emph{Peak-DNA}, a framework for identifying recurrent propagation morphology in global narrative streams. The proposed framework transforms hourly GDELT activity into normalized macro-peak episodes through denoising, macro-peak consolidation, and temporal alignment, and compares propagation morphology using DTW-based similarity analysis. Using GDELT narrative streams, we analyze five high-amplification narrative domains: Iran, Defense, AI, Space/Elon, and Quantum/PQC. The results show that amplification events repeatedly converge toward morphology-level propagation patterns within each narrative domain. DTW-based similarity analysis further demonstrates strong intra-domain structural compactness despite heterogeneous publication dynamics and severe observational noise. These findings suggest a measurable morphology layer beyond semantic content or anomaly intensity, providing an initial structural foundation for recurrence-aware narrative monitoring, propagation family discovery, and longitudinal Narrative Genome analysis.

\keywords{Peak-DNA, Narrative Propagation, Propagation Morphology, Dynamic Time Warping (DTW), Time Series Clustering, GDELT}

\end{abstract}


\section{Introduction}
\label{sec:introduction}

Global narrative systems exhibit explosive amplification around geopolitical conflict, artificial intelligence, defense technology, quantum security transition, and space-related competition. As these large-scale information propagation processes shape public discourse, technological competition, public safety, and national security, understanding narrative propagation has become an important research topic.

Recent studies have leveraged semantic, temporal, and network-level information to investigate narrative diffusion, misinformation, narrative framing, and information cascade prediction~\cite{semantic-classification,forecasting,narrative-modeling}. Although these approaches support analytical tasks such as rumor detection, diffusion forecasting, and narrative classification, comparatively less attention has been devoted to propagation morphology itself as an independent object of structural analysis~\cite{longi-naar-nanlysis}. In particular, it remains largely unexplored whether independently occurring large-scale amplification events exhibit recurrent propagation morphology despite substantial observational noise.

This problem is challenging because global narrative streams are inherently noisy, containing crawler fluctuation, asynchronous publication timing, fragmented burst propagation, temporal jitter, and repeated aftershock amplification~\cite{crawler,async-pub,types}. Consequently, direct comparison of raw propagation waveforms is unreliable, motivating structural representations that remain robust under severe observational noise while preserving recurrent propagation characteristics~\cite{problems}.

To address this challenge, we introduce Peak-DNA, a framework for identifying recurrent propagation morphology in noisy global narrative streams. As described in Section~\ref{sec:framework}, the proposed framework transforms noisy propagation dynamics into normalized macro-peak representations and compares their morphology~\cite{waveform-equiv}. Peak-DNA is introduced as a conceptual abstraction inspired by genomic organization rather than as a literal biological model, providing a structural foundation for recurrence-aware narrative monitoring, propagation family discovery, and future Narrative Genome analysis.

\section{Proposed Framework}
\label{sec:framework}

The proposed Peak-DNA framework identifies recurrent propagation morphology from noisy large-scale narrative streams by progressively transforming raw propagation observations into normalized morphology representations suitable for structural comparison. The framework consists of four sequential stages: noise suppression, macro-peak construction, morphology representation, and Peak-DNA discovery, forming an end-to-end pipeline for structural recurrence analysis.

\paragraph{Noise Suppression}

For each narrative domain $i$, the hourly propagation stream is represented as

\begin{equation}
\mathbf{s}^{(i)}
=
(s^{(i)}_1,\ldots,s^{(i)}_T),
\label{eq:stream}
\end{equation}

where $s_t^{(i)}$ denotes the observed propagation activity at time index $t$.

Real-world narrative propagation is affected by various observational artifacts that obscure local propagation dynamics and hinder reliable structural comparison between independently occurring amplification events. Therefore, the raw propagation stream is transformed into the denoised propagation representation

\begin{equation}
\tilde{\mathbf{s}}^{(i)}
=
\mathcal{F}
\left(
\mathbf{s}^{(i)}
\right),
\label{eq:denoise}
\end{equation}

where $\mathcal{F}(\cdot)$ denotes the smoothing-based denoising operator.
This stage suppresses local stochastic fluctuations while preserving the large-scale propagation structure required for subsequent morphology analysis.

\paragraph{Macro-Peak Construction}

Large-scale narrative amplification is frequently observed as multiple temporally adjacent bursts rather than as a single isolated peak. Treating these fragmented bursts as independent events introduces inconsistent propagation boundaries and hinders reliable morphology comparison across amplification episodes. To obtain a consistent event-level representation, temporally adjacent local peaks are consolidated into unified macro-peaks defined as

\begin{equation}
\mathcal{M}^{(i)}
=
\operatorname{Consolidate}
\left(
\operatorname{Peak}
(
\tilde{\mathbf{s}}^{(i)}
)
\right),
\label{eq:macropeak}
\end{equation}

where $\operatorname{Consolidate}(\cdot)$ merges temporally adjacent local peaks into unified amplification events.

\paragraph{Morphology Representation}

Each macro-peak is transformed into a normalized propagation morphology episode centered on the corresponding amplification event

\begin{equation}
\mathbf{x}^{(i)}_r
=
\Phi
\left(
\tilde{\mathbf{s}}^{(i)},
m_r^{(i)}
\right),
\label{eq:morphology}
\end{equation}

where $\Phi(\cdot,\cdot)$ extracts the propagation pattern surrounding macro-peak $m_r^{(i)}$, resamples each propagation episode into a common temporal representation, and normalizes it within a shared morphology space. This representation suppresses variations caused by propagation magnitude, temporal duration, and temporal offset while preserving the underlying propagation structure, enabling direct structural comparison across independent amplification events. For structural comparison, each morphology episode is represented by its normalized level profile, first-order derivative, and area-normalized profile, providing complementary descriptions of propagation morphology.

The structural similarity between two normalized propagation morphology episodes is evaluated through the morphology distance

\begin{equation}
d^{(i)}_{ab}
=
d_{\mathrm{morph}}
\left(
\mathbf{x}^{(i)}_a,
\mathbf{x}^{(i)}_b
\right),
\label{eq:distance}
\end{equation}

where $d_{\mathrm{morph}}(\cdot,\cdot)$ denotes the DTW-based morphology distance.
Unlike pointwise comparison, DTW aligns temporally shifted propagation dynamics while preserving their structural behavior. The resulting pairwise morphology distances provide the similarity representation required for recurrent morphology discovery.

\paragraph{Peak-DNA Discovery}

The normalized morphology episodes are partitioned into recurrent propagation morphology classes according to their structural similarity

\begin{equation}
\Pi^{(i)}
=
\operatorname{Cluster}
\left(
\{
\mathbf{x}^{(i)}_r
\}
\right),
\label{eq:cluster}
\end{equation}

where $\Pi^{(i)}$ denotes the set of recurrent morphology classes identified within narrative domain $i$.

Among all morphology classes, the dominant Peak-DNA is defined as

\begin{equation}
\mathcal{D}^{(i)*}
=
\arg\max_{\mathcal{C}\in\Pi^{(i)}}
|\mathcal{C}|,
\label{eq:dna}
\end{equation}

where $\mathcal{D}^{(i)*}$ denotes the largest recurrent morphology class.

The dominant recurrence rate is computed as

\begin{equation}
R^{(i)}
=
\frac{
|\mathcal{D}^{(i)*}|
}{
|\mathcal{M}^{(i)}|
},
\label{eq:recurrence}
\end{equation}

where $R^{(i)}$ measures the proportion of macro-peak episodes represented by the dominant Peak-DNA.

\begin{algorithm}[t]
\caption{Peak-DNA Pipeline}
\label{alg:peakdna}

\textbf{Input:}
Narrative domain $i$

\textbf{Output:}
Dominant Peak-DNA $\mathcal{D}^{(i)*}$ and recurrence rate $R^{(i)}$

\begin{enumerate}
\item Construct hourly propagation stream $\mathbf{s}^{(i)}$ using Eq.~\eqref{eq:stream}.
\item Generate denoised propagation stream $\tilde{\mathbf{s}}^{(i)}=\mathcal{F}(\mathbf{s}^{(i)})$ according to Eq.~\eqref{eq:denoise}.
\item Construct macro-peak set $\mathcal{M}^{(i)}$ using Eq.~\eqref{eq:macropeak}.
\item For each $m_r^{(i)}\in\mathcal{M}^{(i)}$, generate normalized morphology episode $\mathbf{x}^{(i)}_r$ using Eq.~\eqref{eq:morphology}.
\item Compute pairwise morphology distance $d^{(i)}_{ab}$ using Eq.~\eqref{eq:distance}.
\item Partition morphology episodes into $\Pi^{(i)}$ using Eq.~\eqref{eq:cluster}.
\item Identify the dominant Peak-DNA $\mathcal{D}^{(i)*}$ using Eq.~\eqref{eq:dna}.
\item Compute dominant recurrence rate $R^{(i)}$ using Eq.~\eqref{eq:recurrence}.
\item Return $\mathcal{D}^{(i)*}$ and $R^{(i)}$.
\end{enumerate}

\end{algorithm}


\section{Experimental Setup}
\label{sec:experimental_setup}

This study analyzes hourly narrative activity streams constructed from GDELT data collected between January~1 and May~12,~2026. The observation period captures repeated large-scale amplification across five high-amplification narrative domains. The analysis focuses on Iran, Defense, AI, Space/Elon, and Quantum/PQC, representing heterogeneous propagation environments associated with geopolitical conflict, defense and security, artificial intelligence, quantum/PQC transition, and space-related technology. Each domain exhibited repeated large-scale propagation bursts, providing suitable conditions for examining recurrent propagation morphology under noisy real-world dynamics. For each narrative domain, hourly propagation streams were processed according to the framework described in Section~\ref{sec:framework} and Algorithm~\ref{alg:peakdna}. The complete processing pipeline, including noise suppression, macro-peak construction, morphology extraction, structural similarity computation, and Peak-DNA discovery, was applied consistently across all narrative domains.

The denoising procedure was performed using Savitzky--Golay smoothing to suppress local stochastic fluctuations while preserving the large-scale propagation morphology. Macro-peaks were identified from the denoised hourly propagation streams, and temporally adjacent peaks were consolidated before morphology extraction so that fragmented bursts originating from the same amplification process were treated as a single propagation event. Each extracted macro-peak episode was resampled to a fixed length and normalized before structural comparison. Structural similarity was computed using a DTW-based structural distance combining Sakoe--Chiba band-constrained DTW over the level profile, first-order derivative, and area-normalized representations of each morphology episode. Peak-DNA classes were identified through agglomerative hierarchical clustering with average linkage. No domain-specific parameter tuning was applied throughout the analysis. Finally, the discovered Peak-DNA classes were summarized using the dominant recurrence measure defined in Eq.~\eqref{eq:recurrence}.

\section{Results and Discussion} 
\label{sec:results}

\subsection{Macro-Peak Emergence Across Narrative Domains}

\begin{figure*}[t]
\centering
\includegraphics[width=0.87\linewidth]{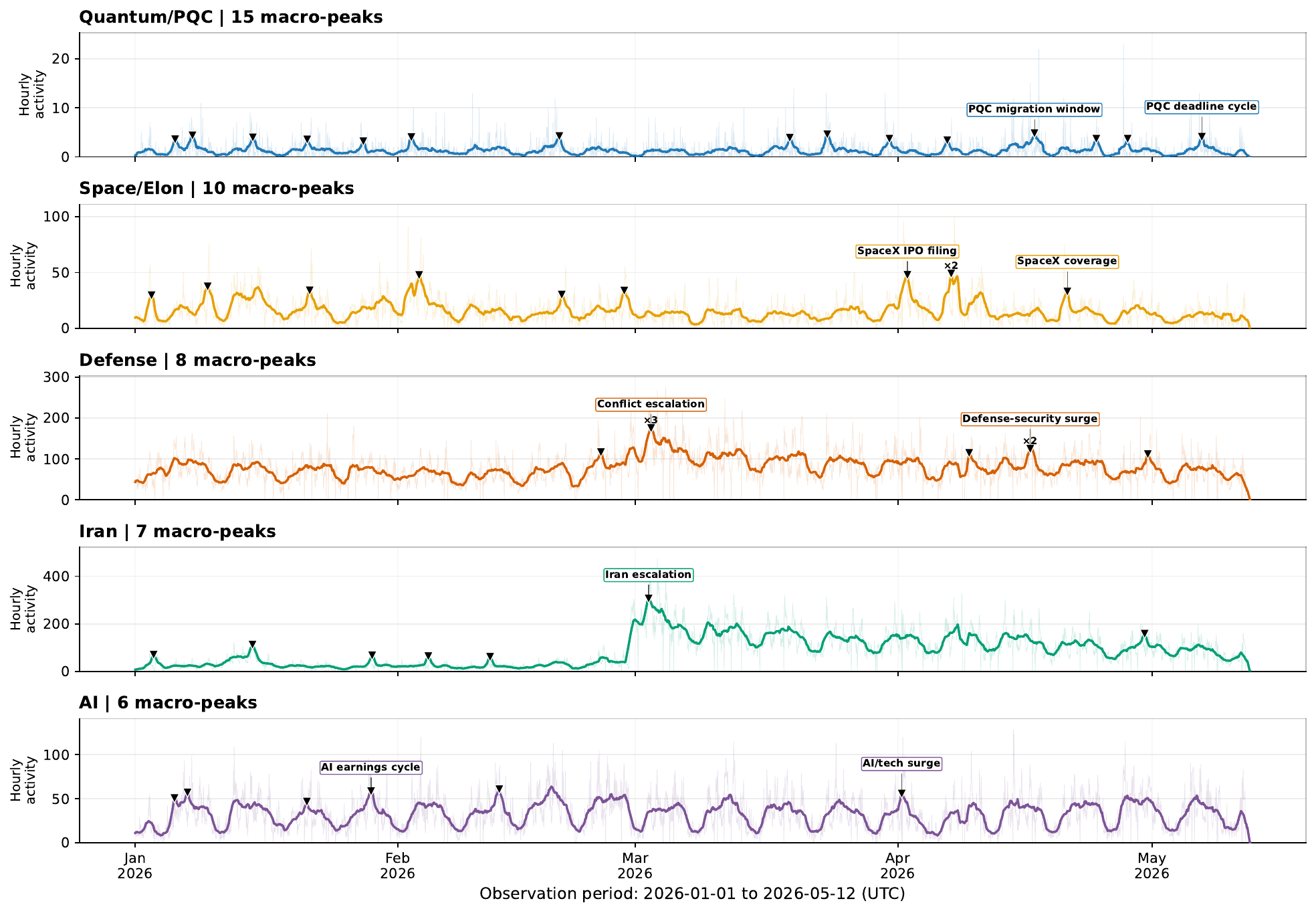}
\caption{
Hourly narrative propagation streams and consolidated macro-peak events across the selected narrative domains.
}
\label{fig:macropeak_streams}
\end{figure*}

The first analysis investigates whether large-scale narrative streams exhibit recurring amplification episodes suitable for morphology-level analysis. Across the observation period, the five narrative domains repeatedly exhibited propagation bursts that were identified as macro-peak events. Figure~\ref{fig:macropeak_streams} presents the hourly propagation streams together with the corresponding macro-peak events identified in each narrative domain. Across the five narrative domains, the proposed framework identified 15 macro-peak episodes for Quantum/PQC, 10 for Space/Elon, 8 for Defense, 7 for Iran, and 6 for AI. The repeated emergence of macro-peak episodes across all narrative domains provides event-level observations for the subsequent morphology-level recurrence analysis.

\subsection{Peak-DNA Recurrence}

\begin{figure*}[t]
\centering
\includegraphics[width=0.87\linewidth]{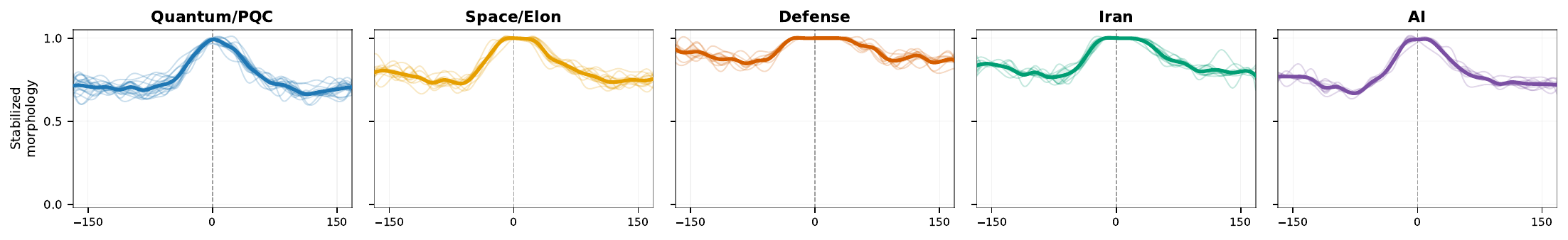}
\caption{
Peak-centered recurrence of normalized propagation morphology across the selected narrative domains.
}
\label{fig:dna_waveforms}
\end{figure*}

The second analysis investigates whether consolidated macro-peak episodes exhibit recurrent propagation morphology. Figure~\ref{fig:dna_waveforms} presents normalized morphology episodes aligned at the macro-peak center across the selected narrative domains. Each transparent trajectory represents an individual macro-peak episode, while the thick curve represents the dominant Peak-DNA morphology. The results show that independent macro-peak episodes exhibit recurrent propagation morphology within each analyzed narrative domain. Although the dominant morphology differs across narrative domains, individual episodes remain tightly clustered around the corresponding Peak-DNA, exhibiting consistent rise--peak--decay behavior with recurrent rebound-aftershock dynamics. These observations support the central hypothesis that independently occurring large-scale amplification events exhibit recurrent propagation morphology.

\subsection{Morphology-Level Phenotype Analysis}

\begin{figure*}[t]
\centering
\includegraphics[width=1\linewidth]{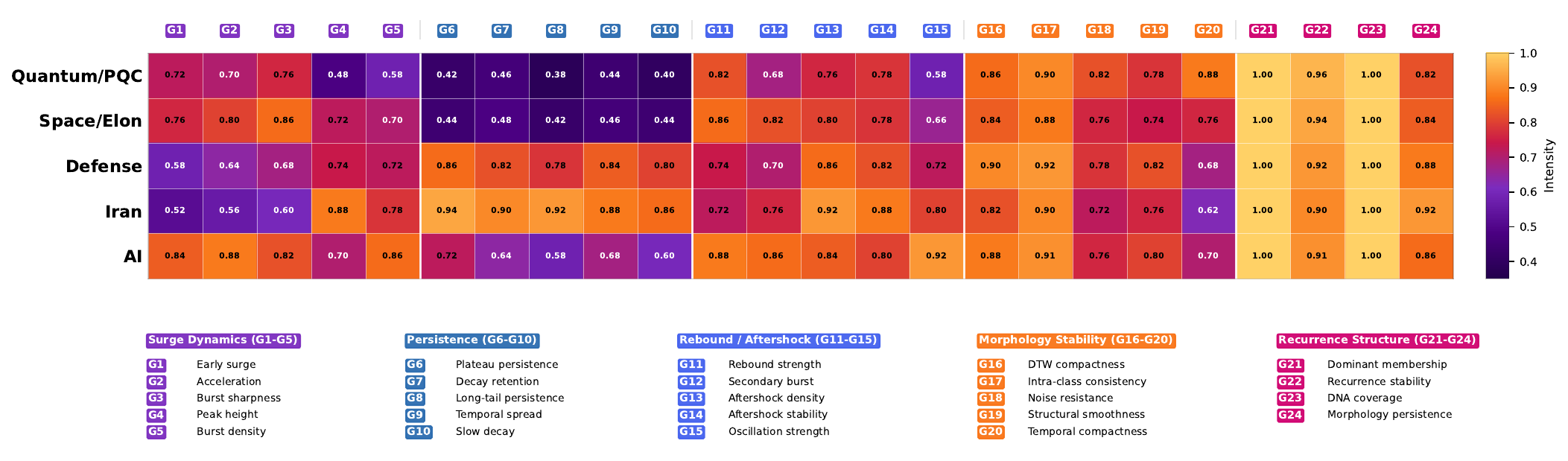}
\caption{
Phenotype profiles of Peak-DNA across the selected narrative domains.
}
\label{fig:dna_traitmap}
\end{figure*}

The third analysis investigates whether the discovered Peak-DNA structures can be interpreted as morphology-level propagation phenotypes rather than single waveform templates. Figure~\ref{fig:dna_traitmap} presents the phenotype profiles of Peak-DNA across the selected narrative domains, where 24 morphology genes are organized into five phenotype categories: surge dynamics, persistence, rebound-aftershock behavior, morphology stability, and recurrence structure. The results show that Peak-DNA can be represented through a consistent genome-like phenotype profile that distinguishes domain-specific recurrent morphology beyond a single propagation waveform. 

\subsection{Structural Validation}

\begin{figure*}[t]
\centering
\includegraphics[width=0.7\linewidth]{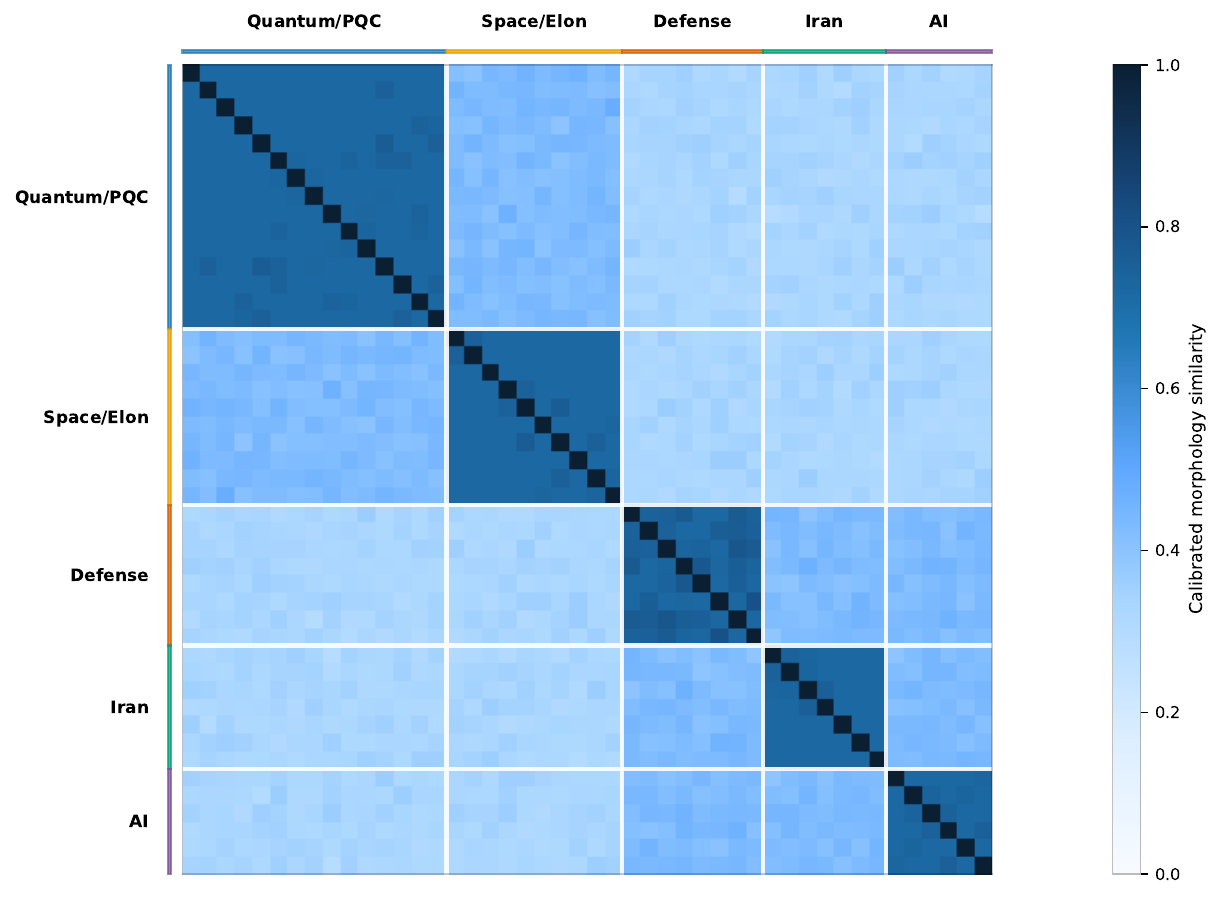}
\caption{
Pairwise DTW-based morphology similarity across normalized macro-peak propagation episodes.
}
\label{fig:dtw_similarity}
\end{figure*}

To further examine the structural relationships among the discovered Peak-DNA patterns, Figure~\ref{fig:dtw_similarity} presents pairwise morphology similarity across normalized macro-peak propagation episodes. The within-domain regions indicate high structural similarity, confirming the recurrence of dominant propagation morphology. Beyond the strong intra-domain similarity, several narrative domains also exhibit partially shared morphology patterns. Two visually coherent higher-level groups are observable: Quantum/PQC--Space/Elon and Defense--Iran--AI. Although these groupings remain exploratory, they suggest that semantically distinct narrative domains may share broader propagation morphology beyond their individual domain-specific Peak-DNA patterns. Collectively, the waveform recurrence, phenotype profiles, and DTW-based structural similarity support the morphology representation defined in Section~\ref{sec:framework}. These observations further suggest that recurrent propagation morphology may be organized into higher-level morphology families, providing an initial structural basis for future Narrative Genome analysis. The present study establishes an initial structural baseline using five high-amplification narrative domains within a single observation period. Future work will extend the proposed framework toward longitudinal recurrence analysis, morphology family evolution, and Narrative Genome mapping.

\section{Conclusion}

We proposed Peak-DNA, a morphology-oriented framework for identifying recurrent propagation structure in noisy global narrative streams. The proposed framework consolidates fragmented amplification bursts into normalized macro-peak representations for structural morphology comparison. Using hourly GDELT streams, five high-amplification narrative domains consistently exhibited dominant Peak-DNA recurrence after macro-peak consolidation and morphology normalization. The waveform recurrence, phenotype profiles, and DTW-based structural similarity jointly demonstrate repeatable morphology-level structural patterns beyond semantic content or anomaly intensity. These findings provide an initial foundation for recurrence-aware narrative monitoring, propagation family discovery, morphology evolution tracking, and future Narrative Genome analysis.


\begin{credits}

\subsubsection{\ackname}
The authors gratefully acknowledge Prof. Hiroaki Takada and Prof. Yutaka Matsubara of the Graduate School of Informatics, Nagoya University, for their valuable academic support and research guidance.

\subsubsection{\discintname}
The authors declare that there are no competing interests relevant to the content of this article.

\end{credits}

\end{document}